\documentclass[10pt,twocolumn]{article}
\usepackage[letterpaper, margin=0.75in, columnsep=0.3in]{geometry}
\usepackage{times}
\usepackage[T1]{fontenc}
\usepackage{graphicx}
\usepackage{booktabs}
\usepackage{multirow}
\usepackage{xcolor}
\usepackage{colortbl}
\usepackage{tabularx}
\usepackage{amsmath}
\usepackage[hidelinks]{hyperref}
\usepackage[font=small,labelfont=bf]{caption}
\usepackage{float}
\usepackage{enumitem}
\setlist{nosep, leftmargin=*}
\usepackage{titlesec}
\titleformat{\section}{\large\bfseries}{\thesection}{0.6em}{}
\titleformat{\subsection}{\normalsize\bfseries}{\thesubsection}{0.5em}{}
\titleformat{\subsubsection}{\normalsize\itshape}{\thesubsubsection}{0.5em}{}
\titlespacing{\section}{0pt}{8pt plus 2pt minus 2pt}{3pt plus 1pt minus 1pt}
\titlespacing{\subsection}{0pt}{6pt plus 2pt minus 2pt}{2pt plus 1pt minus 1pt}
\renewenvironment{abstract}{\noindent\textbf{Abstract}\par}{\par\vspace{6pt}}
\begin{document}

\twocolumn[{%
\begin{center}
{\LARGE\bfseries Beyond Benchmarks: How Users Evaluate AI Chat Assistants\par}
\vspace{14pt}
\begin{tabular}{ccc}
\begin{minipage}[t]{0.30\textwidth}\centering
{\large Moiz Sadiq Awan}\\[2pt]
{\normalsize Independent Researcher}\\[1pt]
{\small \texttt{moizsawan@gmail.com}}
\end{minipage} &
\begin{minipage}[t]{0.30\textwidth}\centering
{\large Muhammad Salman Munaf}\\[2pt]
{\normalsize Independent Researcher}\\[1pt]
{\small \texttt{salmanmunaf96@gmail.com}}
\end{minipage} &
\begin{minipage}[t]{0.30\textwidth}\centering
{\large Muhammad Haris Noor}\\[2pt]
{\normalsize Independent Researcher}\\[1pt]
{\small \texttt{harisnoor674@gmail.com}}
\end{minipage}
\end{tabular}
\vspace{12pt}
\end{center}
}]

\begin{abstract}
Automated benchmarks dominate the evaluation of large language models, yet no systematic study has compared user satisfaction, adoption motivations, and frustrations across competing platforms using a consistent instrument. We address this gap with a cross-platform survey of 388 active AI chat users, comparing satisfaction, adoption drivers, use case performance, and qualitative frustrations across seven major platforms: ChatGPT, Claude, Gemini, DeepSeek, Grok, Mistral, and Llama. Three broad findings emerge. First, the top three platforms (Claude, ChatGPT, and DeepSeek) receive statistically indistinguishable satisfaction ratings despite vast differences in funding, team size, and benchmark performance. Second, users treat these tools as interchangeable utilities rather than sticky ecosystems: over 80\% use two or more platforms, and switching costs are negligible. Third, each platform attracts users for different reasons: ChatGPT for its interface, Claude for answer quality, DeepSeek through word-of-mouth, and Grok for its content policy, suggesting that specialization, not generalist dominance, sustains competition. Hallucination and content filtering remain the most common frustrations across all platforms. These findings offer an early empirical baseline for a market that benchmarks alone cannot characterize, and point toward competitive plurality rather than winner-take-all consolidation among engaged users.
\end{abstract}

\vspace{6pt}

\section{Introduction}

Three years after ChatGPT reached 100 million users in record time~\cite{hu2023}, the AI chat assistant market has not followed the trajectory that early predictions suggested. Rather than a single dominant platform, users now choose among at least seven major competitors: OpenAI's ChatGPT, Anthropic's Claude, Google's Gemini, Meta's Llama, xAI's Grok, DeepSeek, and Mistral's Le Chat. Hundreds of millions of people interact with these systems daily~\cite{similarweb2024}. Yet for all the capital, compute, and engineering talent flowing into this market, a basic empirical question remains open: what do users think of these platforms, and why do they choose one over another?

Despite the enormous scale of adoption, empirical understanding of user preferences has not kept pace. The AI research community has developed sophisticated automated benchmarks: MMLU~\cite{hendrycks2021} assesses multi-task understanding across 57 academic subjects, HumanEval~\cite{chen2021} evaluates code generation, and the LMSYS Chatbot Arena~\cite{chiang2024} provides crowd-sourced pairwise preference rankings in controlled settings. These benchmarks are useful for tracking technical progress, but they leave critical questions unanswered. What factors beyond raw capability drive adoption and retention? How do users allocate their usage across competing platforms? What pain points persist despite rapid technical improvement?

These questions have practical consequences. For product teams, knowing which attributes users actually value can redirect development effort away from benchmark-chasing toward features that affect daily experience. For market analysts, the degree of platform switching and multi-homing determines whether the AI assistant market is consolidating or sustaining genuine competition. For regulators, user satisfaction data bear on questions of market concentration and consumer welfare. For evaluation researchers, understanding what drives real-world platform choice can inform the design of more ecologically valid assessment protocols. Yet companies continue to invest billions in model development guided primarily by benchmark rankings, and policy discussions about AI market dominance have proceeded with little evidence about how consumers actually experience competing products.

This paper presents results from a cross-platform survey ($N=388$) conducted in late 2025, examining preferences across seven major AI chat platforms. While existing user studies have focused on individual platforms (predominantly ChatGPT), comparative data across multiple competing platforms remains scarce. Because the sample is drawn from technology-oriented communities, our findings characterize the preferences of active, engaged AI chat users rather than the general population. Our contributions are as follows:

\begin{itemize}
\item We document pervasive multi-platform usage and ChatGPT-centric switching patterns, providing systematic evidence that AI chat tools function as interchangeable utilities rather than sticky ecosystems.
\item We establish empirical baselines for user satisfaction across seven platforms, finding reported satisfaction parity among the top three (Claude, ChatGPT, and DeepSeek) despite vast resource asymmetries.
\item We identify distinct adoption profiles and domain specializations that differentiate each platform's competitive positioning, suggesting that specialization rather than generalist dominance sustains competition.
\item We uncover a large first-mover anchoring effect: ChatGPT users who adopted it as their first AI tool rate satisfaction 1.34 points higher than those who arrived from a competitor, highlighting the role of habit and default effects.
\item We present a thematic analysis of 329 open-ended responses identifying hallucination and content moderation as the two most persistent user frustrations, representing a direct engineering tradeoff.
\end{itemize}

\section{Related Work}

\subsection{Automated LLM Evaluation}

The evaluation of large language models has been dominated by automated benchmarks targeting specific capabilities. MMLU~\cite{hendrycks2021} tests multi-task understanding across 57 academic subjects, providing a broad measure of factual knowledge and reasoning. HumanEval~\cite{chen2021} and MBPP~\cite{austin2021} evaluate functional code generation by testing whether model outputs pass unit tests. GSM8K~\cite{cobbe2021} tests mathematical reasoning through grade-school word problems, while TruthfulQA~\cite{lin2022} probes factual accuracy and resistance to common misconceptions. More comprehensive suites such as HELM~\cite{liang2023} and BIG-Bench~\cite{srivastava2023} attempt broad-coverage assessment across dozens of tasks simultaneously, though they remain bounded by what can be automatically scored.

A significant advance in ecological validity came with the LMSYS Chatbot Arena~\cite{chiang2024}, which collects human preference judgments through blind pairwise comparisons of model outputs in real conversational settings, producing Elo-style rankings that have become a widely cited industry reference. Zheng et al.~\cite{zheng2023} further proposed MT-Bench, a multi-turn benchmark using strong LLMs as judges, which correlates well with human preferences but still operates at the response level rather than the platform experience level.

However, both Arena and MT-Bench capture isolated response quality in single-turn or short multi-turn exchanges. They do not measure the broader user experience that encompasses interface design, pricing, ecosystem integration, content policy, reliability over extended use, and the accumulated satisfaction that drives real-world platform choice. A shared limitation is that these approaches treat model quality as a single scalar, when in practice users weigh multiple factors simultaneously (accuracy, speed, style, policy permissiveness, and cost), and these factors interact in complex, context-dependent ways.

\subsection{Technology Adoption and Platform Economics}

The Technology Acceptance Model (TAM; Davis~\cite{davis1989}) and its successor UTAUT~\cite{venkatesh2003} identify perceived usefulness, ease of use, social influence, and facilitating conditions as key determinants of technology adoption. The Expectation-Confirmation Model~\cite{bhattacherjee2001} extends this framework to continued usage, demonstrating that post-adoption satisfaction depends on the gap between experienced performance and initial expectations. In the context of AI tools, Jo~\cite{jo2024} adapted TAM to study ChatGPT adoption among South Korean users, finding that output quality and ease of use were the strongest predictors of continued use intention.

The concept of multi-homing, in which consumers simultaneously maintain active usage across competing platforms, has been studied extensively in platform economics. Rochet and Tirole~\cite{rochet2003} and Armstrong~\cite{armstrong2006} established foundational models of two-sided market competition, showing that differentiation and switching costs jointly determine market structure. Shapiro and Varian~\cite{shapiro1998} argued that in information goods markets, network effects and switching costs typically drive toward concentration. However, Bommasani et al.~\cite{bommasani2022} noted that foundation model markets may follow different dynamics due to low switching costs, minimal network effects, and rapid capability convergence, a theoretical prediction that our data allow us to examine empirically.

Cennamo and Santalo~\cite{cennamo2013} showed that platform differentiation strategies vary by market maturity: early-stage markets compete on breadth, while mature markets compete on specialized excellence. This framework is relevant to the AI assistant landscape, where we observe a similar transition from generalist competition toward domain-specific positioning.

\subsection{User Studies of AI Chat Tools}

Empirical user studies of AI chat tools have grown rapidly but remain limited in scope. Skjuve et al.~\cite{skjuve2023} conducted qualitative interviews with 18 users of the Replika companion chatbot, finding that human-chatbot relationships develop through stages of curiosity, self-disclosure, and sustained engagement. While Replika differs from the task-oriented assistants we study, this work highlights the relational dimensions of human-AI interaction that satisfaction surveys alone may miss. Ray~\cite{ray2023} provided a qualitative assessment of ChatGPT across professional domains, cataloging strengths and limitations but without comparative data from alternative platforms. Jansen et al.~\cite{jansen2023} analyzed user interactions with ChatGPT through conversation logs, finding that query complexity and topic diversity increased over time as users developed more sophisticated prompting strategies.

In educational contexts, Kasneci et al.~\cite{kasneci2023} reviewed opportunities and challenges of LLMs for teaching and learning, while Baidoo-Anu and Ansah~\cite{baidoo2023} examined student perceptions of ChatGPT as a learning aid. In professional settings, Noy and Zhang~\cite{noy2023} conducted a randomized experiment showing that ChatGPT substantially increased writing productivity, with the largest gains among lower-ability workers. Brynjolfsson et al.~\cite{brynjolfsson2023} found similar productivity effects for customer service agents using AI assistance.

More recently, a Pew Research Center survey conducted in January 2025 found that roughly half of U.S.\ adults had used an LLM, with multi-platform usage common among adopters~\cite{pew2025}. A Rethink Priorities survey of nearly 2,000 U.S.\ tech workers in early 2025 reported 91\% LLM usage, with ChatGPT dominant at 82\% but substantial uptake of Claude (16\%) and Gemini (29\%)~\cite{rethink2025}. And in the enterprise segment, a mid-2025 Menlo Ventures survey of 150 technical leaders found that Anthropic had overtaken OpenAI in enterprise market share, with most organizations deploying multiple models simultaneously~\cite{menlo2025}. These industry surveys confirm the multi-homing patterns our study investigates at the individual user level.

What remains missing, however, is a study that compares user satisfaction, adoption motivations, and qualitative frustrations across multiple platforms using a single, consistent instrument. The studies reviewed above each examine a single platform (overwhelmingly ChatGPT) in isolation or report aggregate market statistics without per-platform satisfaction data. Our study addresses this gap by surveying users of seven platforms with identical evaluation instruments, enabling within-subjects comparison and cross-platform analysis.

\section{Methodology}

\subsection{Survey Design and Instrument}

The survey instrument was developed in Qualtrics and organized into four sections.

\textbf{Section 1: Demographics and Usage Profile.} This section captured respondent occupation (8 categories with free-text ``Other''), country of residence, and overall AI chat usage frequency (daily, weekly, monthly, or less).

\textbf{Section 2: Model Selection.} Respondents identified all AI chat models they had actively used in the past six months from a checklist. The initial instrument included ChatGPT, Claude, Gemini, Llama, and Grok with an ``Other'' option. Analysis of early responses ($n \approx 80$) revealed substantial DeepSeek and Mistral usage via ``Other,'' prompting the addition of these as explicit options to the live survey. This mid-survey instrument change warrants transparent reporting. The modification was made during week 2 of data collection (approximately October 21, 2025). Early respondents could and did report DeepSeek and Mistral usage through the free-text ``Other'' field; 38 DeepSeek users and 18 Mistral users appear in the first half of the collection period, compared to 20 and 12 respectively in the second half. We found no significant difference in ChatGPT satisfaction between early and late respondents ($M=3.83$ vs.\ $M=3.62$, Mann-Whitney $p=0.524$), suggesting the change did not alter response patterns. Nonetheless, analyses involving DeepSeek and Mistral should be interpreted with this instrument modification in mind.

\textbf{Section 3: Per-Model Evaluation.} For each model the respondent reported using, an identical evaluation block was administered. This within-subjects design enables direct cross-platform comparison while controlling for individual differences. Each block contained:

\begin{itemize}
\item Overall satisfaction on a 5-point Likert scale (1=Extremely dissatisfied to 5=Extremely satisfied)
\item Nine adoption drivers rated on a 5-point importance scale: perceived value for money, answer quality, multimodal capability, UI/UX design, response speed, work-task suitability, word-of-mouth recommendation, promotional discount, and censorship/content policy alignment
\item Six use case performance ratings on a 5-point Likert scale: content creation, communication assistance, learning and research, technical and analytical tasks, productivity and automation, and business and professional use
\item Current subscription plan (free, individual paid, team/enterprise, unsure)
\item Response to a hypothetical 25\% price increase (keep current plan, downgrade, switch to competitor, stop using AI tools)
\item Whether this was the respondent's first AI chat tool (yes/no)
\item Usage tenure (less than 3 months, 3--6 months, 6--12 months, 12--18 months, more than 18 months)
\item Switching history (whether they switched from another primary model, and if so, which one)
\end{itemize}

\textbf{Section 4: Open-Ended Responses.} Two free-text questions elicited qualitative data: one asking about the most frustrating limitation of AI chat models, and another asking what feature respondents most wished their preferred model had or did better. These questions were presented after all quantitative blocks to avoid framing effects.

\subsection{Sample and Distribution}

The survey was distributed through technology-focused online communities (primarily Reddit subreddits related to AI, machine learning, and productivity tools) and professional networks during late 2025, employing convenience sampling. This distribution strategy was selected to reach active AI chat users across multiple platforms, as no comprehensive sampling frame exists for the population of interest.

The survey was administered in two waves using the same Qualtrics instrument. The first wave ($n=151$) used the original five-model checklist (ChatGPT, Claude, Gemini, Llama, Grok) plus a free-text ``Other'' field. The second wave ($n=237$), launched after DeepSeek and Mistral were added as explicit options, constitutes the primary analytic sample for all quantitative per-model comparisons. Across both waves, the survey yielded 388 total responses, of which 262 provided usable demographic and model selection data and 180 fully completed all sections. All quantitative per-model analyses (satisfaction, adoption drivers, use case ratings, subscription data, switching behavior) draw on the second-wave sample ($N=237$; demographic subsample $n=171$; platform-level $n$ values reported per analysis). Qualitative responses from both waves are combined ($n=329$). Sample sizes are reported per analysis throughout to ensure transparency.

\begin{table}[t]
\centering
\caption{Respondent Demographics (second-wave sample, $n=171$ with complete data).}
\label{tab:demographics}
\small
\begin{tabular}{lrr}
\toprule
\textbf{Characteristic} & \textbf{n} & \textbf{\%} \\
\midrule
\textit{Occupation} & & \\
\quad Developer / Software Engineer & 48 & 28.1 \\
\quad Student & 38 & 22.2 \\
\quad Other & 32 & 18.7 \\
\quad Researcher / Academic & 18 & 10.5 \\
\quad Consultant / Business Prof. & 16 & 9.4 \\
\quad Data Analyst / Data Scientist & 11 & 6.4 \\
\quad Writer / Content Creator & 4 & 2.3 \\
\quad Medical Professional & 4 & 2.3 \\
\midrule
\textit{Usage Frequency} & & \\
\quad Daily or almost daily & 136 & 79.5 \\
\quad Weekly & 23 & 13.5 \\
\quad Monthly or less & 12 & 7.0 \\
\midrule
\textit{Geographic Region ($n=165$)} & & \\
\quad North America & 63 & 38.2 \\
\quad South Asia & 60 & 36.4 \\
\quad Europe (9 countries) & 29 & 17.6 \\
\quad Asia-Pacific (5 countries) & 8 & 4.8 \\
\quad Middle East \& Latin America & 5 & 3.0 \\
\bottomrule
\end{tabular}
\vspace{2pt}
\\\footnotesize\textit{Note:} Respondents spanned 37 countries across five continents.
\end{table}

The resulting sample (Table~\ref{tab:demographics}) is composed primarily of technology professionals and students, with 79.5\% reporting daily AI chat usage. Respondents represented 37 countries, with the largest concentrations in North America (38.2\%), South Asia (36.4\%), and Europe (17.6\%). This profile likely overrepresents power users and technology professionals, a limitation we address in Section~\ref{sec:limitations}.

\subsection{Data Cleaning and Quality Assurance}

Responses were screened for quality using three criteria: (1) minimum completion time of 3 minutes to exclude random clicking, (2) non-contradictory model selection (e.g., respondents who selected models but then skipped all evaluation blocks were excluded), and (3) non-identical responses across all Likert items within a model block (straight-lining detection). Responses failing these criteria were removed, yielding the final analytic sample.

\subsection{Analytical Methods}

Given ordinal Likert data (responses on a fixed scale from 1 to 5, where the intervals between points are not necessarily equal), we employ non-parametric tests throughout, as these do not assume a normal distribution. Kruskal-Wallis $H$ tests assess whether satisfaction differs significantly across platforms as a group (analogous to a one-way ANOVA for ranked data); post-hoc pairwise Mann-Whitney $U$ tests then identify which specific platform pairs differ, with Bonferroni correction (adjusted $\alpha = 0.0024$ for 21 pairwise comparisons among 7 platforms) to control for the increased risk of false positives when making multiple comparisons. Effect sizes are reported as $\varepsilon^2$ for Kruskal-Wallis tests (small $\geq 0.01$, medium $\geq 0.06$, large $\geq 0.14$), Cohen's $d$ for pairwise comparisons, and Cram\'{e}r's $V$ for chi-square tests. All confidence intervals are 95\% unless noted. Spearman rank correlations assess associations between ordinal variables.

\begin{figure}[t]
\centering
\includegraphics[width=\columnwidth]{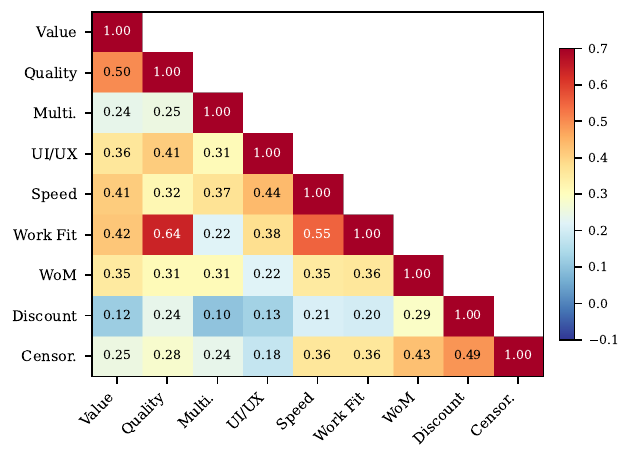}
\caption{Inter-item correlation matrix for the nine-item adoption driver scale (ChatGPT, $n=139$). Two clusters are visible: product quality items (upper-left) and external factor items (lower-right).}
\label{fig:corr}
\end{figure}

\textbf{Scale reliability.} Internal consistency was assessed via Cronbach's $\alpha$ for both the six-item use case performance scale and the nine-item adoption driver scale. The use case scale showed good reliability across platforms: ChatGPT $\alpha=0.80$ ($n=137$), Claude $\alpha=0.79$ ($n=42$), Gemini $\alpha=0.82$ ($n=88$), DeepSeek $\alpha=0.85$ ($n=41$). The nine-item adoption driver scale showed comparable reliability: ChatGPT $\alpha=0.80$ ($n=139$), Claude $\alpha=0.75$ ($n=45$), Gemini $\alpha=0.79$ ($n=89$), DeepSeek $\alpha=0.80$ ($n=41$). The mean inter-item correlation for the ChatGPT adoption driver scale was $r=0.32$, within the recommended range of 0.15--0.50 for broad constructs. Inspection of the inter-item correlation matrix (Figure~\ref{fig:corr}) revealed two interpretable clusters: a ``product quality'' cluster (Quality, Value, Speed, Work Fit, UI/UX; mean within-cluster $r=0.45$) and an ``external factors'' cluster (Word-of-Mouth, Censorship, Discount; mean within-cluster $r=0.40$), with weaker cross-cluster correlations ($r=0.20$--$0.35$). We report these descriptively rather than conducting formal factor analysis given sample size constraints.

Open-ended responses ($n=329$) were analyzed through keyword-assisted thematic coding: initial codes were generated inductively from the first 50 responses, then applied systematically to the full corpus with iterative refinement.

\subsection{Robustness Checks}

To assess whether partial survey completion introduced systematic bias, we compared full completers ($n=129$, all sections finished) with partial responders ($n=108$, at least one model block completed) on key variables. Full completers were more likely to report daily usage (81.4\% vs.\ 28.7\%), consistent with engaged users being more willing to complete a lengthy survey. However, ChatGPT satisfaction did not differ significantly between the two groups (full $M=3.83$, $n=119$; partial $M=3.50$, $n=20$; Mann-Whitney $U=1352$, $p=0.301$). Claude and Gemini partial-responder subsamples were too small ($n<6$) for meaningful comparison.

We also compared early-half ($n=119$) and late-half ($n=118$) respondents, split at the collection midpoint. ChatGPT satisfaction was stable across collection periods (early $M=3.83$ vs.\ late $M=3.62$, $p=0.524$). DeepSeek and Mistral users appeared in both halves (early: 38 DeepSeek, 18 Mistral; late: 20 DeepSeek, 12 Mistral), indicating that the mid-survey addition of these options did not create a sharp discontinuity. These checks suggest that the main satisfaction findings are not artifacts of differential completion or temporal effects, though the limited statistical power of these comparisons warrants caution.

\section{Results}

\subsection{Market Structure: Adoption, Multi-Homing, and Switching}

\begin{figure}[t]
\centering
\includegraphics[width=\columnwidth]{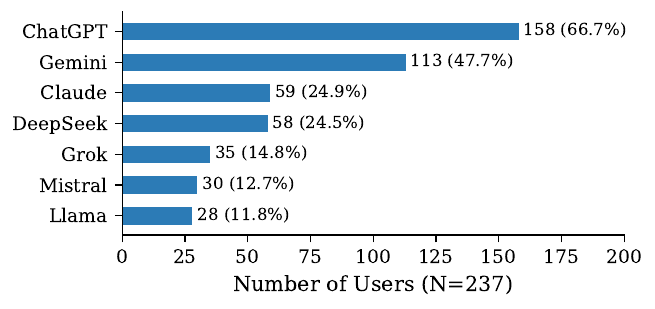}
\caption{Self-reported AI model usage in the past six months ($N=237$). Percentages reflect the share of all respondents reporting use.}
\label{fig:usage}
\end{figure}

ChatGPT was the most widely used platform, reported by 158 of 237 respondents (66.7\%), followed by Gemini (113, 47.7\%), Claude (59, 24.9\%), DeepSeek (58, 24.5\%), Grok (35, 14.8\%), Mistral (30, 12.7\%), and Llama (28, 11.8\%) (Figure~\ref{fig:usage}). The dominance of ChatGPT in raw adoption is consistent with its first-mover advantage and brand recognition, though its lead over Gemini is smaller than market share estimates from web traffic data might suggest~\cite{similarweb2024}, likely reflecting the tech-savvy composition of our sample.

\begin{figure}[t]
\centering
\includegraphics[width=0.85\columnwidth]{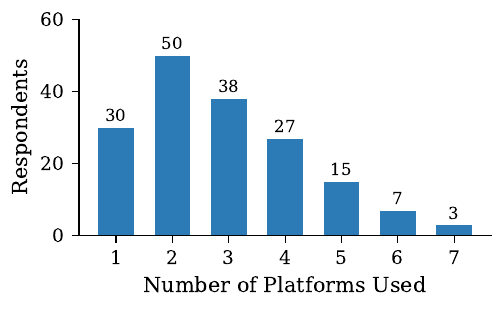}
\caption{Distribution of the number of AI chat platforms used per respondent ($n=170$). The mean is 2.83 and the median is 3.}
\label{fig:multimodel}
\end{figure}

Multi-platform usage is widespread. Among 170 active users providing complete data, the mean number of platforms used was 2.83 ($SD = 1.49$, median = 3). A substantial 82.4\% reported using two or more models, and 52.9\% used three or more (Figure~\ref{fig:multimodel}). Only 17.6\% were single-platform users. This pattern points to low switching costs and task-specific model selection rather than single-ecosystem commitment. The distribution is right-skewed, with 7.1\% of respondents using six or seven platforms simultaneously.

\noindent\textbf{Key takeaway:} The AI chat market is not consolidating around a single winner. Users routinely maintain portfolios of two to three platforms, suggesting low switching costs and task-specific model selection.

\begin{figure}[t]
\centering
\includegraphics[width=\columnwidth]{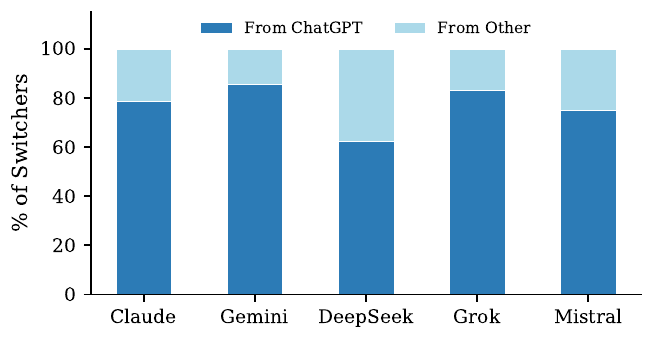}
\caption{Source of platform switchers, showing the dominance of ChatGPT as the origin platform.}
\label{fig:switching}
\end{figure}

ChatGPT functions as the primary source from which users explore and migrate to alternative platforms. Of the 14 respondents who switched to Claude as their primary tool, 11 (78.6\%) came from ChatGPT. Similarly, 18/21 Gemini switchers (85.7\%), 5/6 Grok switchers (83.3\%), and 6/8 Mistral switchers (75.0\%) previously used ChatGPT as their primary model (Figure~\ref{fig:switching}). DeepSeek was the partial exception: while 50.0\% of its switchers came from ChatGPT, 50.0\% came from other platforms (Claude and Gemini), reflecting its appeal to users already exploring beyond the dominant platform.

Mistral had the highest switching-in rate (42.1\%, 8/19 of all Mistral users), followed by Claude (31.1\%, 14/45). ChatGPT had the lowest switching-in rate (8.6\%, 12/139), consistent with its position as the default starting point rather than a destination for switchers.

Among Claude users, the number of models concurrently used was negatively correlated with Claude satisfaction (Spearman $\rho=-0.39$, $p=0.008$, $n=45$), suggesting that extensive multi-homing may indicate ongoing search for a better alternative rather than complementary use. This pattern was not significant for ChatGPT ($\rho=-0.15$, $p=0.073$) or Gemini ($\rho=-0.07$, $p=0.534$), possibly because these platforms serve as default or generalist choices that persist regardless of satisfaction level.

\noindent\textbf{Key takeaway:} ChatGPT is the dominant origin point for platform switchers. Users who leave ChatGPT distribute across competitors, while DeepSeek uniquely draws users from beyond the ChatGPT ecosystem.

\subsection{Satisfaction Convergence Across Platforms}

\begin{figure}[t]
\centering
\includegraphics[width=\columnwidth]{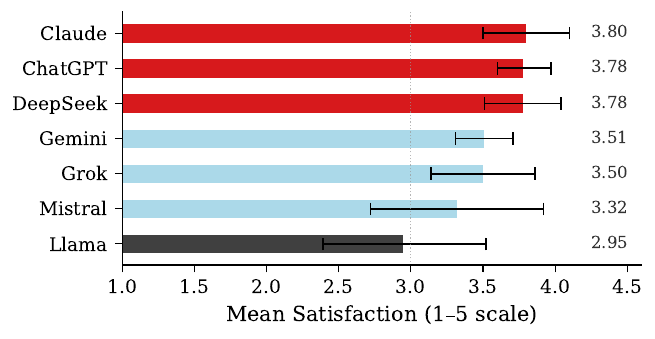}
\caption{Mean satisfaction by platform with 95\% confidence intervals. The dotted line indicates the neutral midpoint (3.0).}
\label{fig:satisfaction}
\end{figure}

\begin{figure}[t]
\centering
\includegraphics[width=\columnwidth]{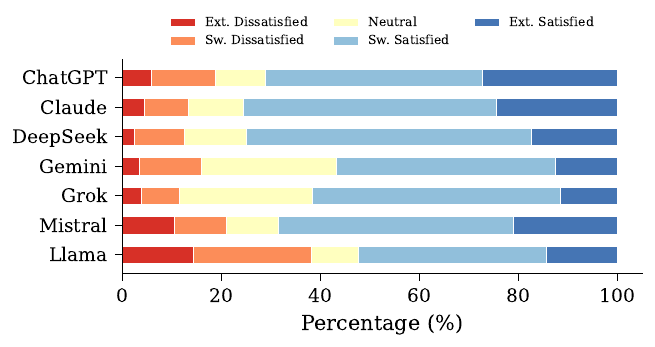}
\caption{Full satisfaction distribution by platform. The stacked bars show the percentage of respondents at each satisfaction level.}
\label{fig:satdist}
\end{figure}

Given this multi-platform landscape, a natural question is whether users perceive meaningful quality differences across platforms. A Kruskal-Wallis test revealed a significant omnibus difference in satisfaction across the seven platforms ($H=16.39$, $df=6$, $p=0.012$, $\varepsilon^2=0.043$). The effect size is small by conventional thresholds, indicating that while the overall distribution of satisfaction differs across platforms, the magnitude of this difference is modest. The top three platforms had broadly overlapping confidence intervals: Claude ($M=3.80$, $SD=1.01$, 95\% CI [3.50, 4.10], $n=45$), ChatGPT ($M=3.78$, $SD=1.10$, [3.60, 3.97], $n=139$), and DeepSeek ($M=3.78$, $SD=0.86$, [3.51, 4.04], $n=40$), with no significant pairwise differences among them (all $p > 0.50$). Llama scored lowest ($M=2.95$, $SD=1.32$, [2.39, 3.52], $n=21$), and was the only platform below the neutral midpoint of 3.0. However, both Llama ($n=21$) and Mistral ($n=19$) have sample sizes too small for reliable inference; findings involving these platforms should be treated as descriptive rather than inferential. Figure~\ref{fig:satisfaction} displays means with 95\% confidence intervals; Figure~\ref{fig:satdist} shows the full distribution.

Post-hoc Mann-Whitney tests with Bonferroni correction found no individual comparison reaching the adjusted significance threshold ($\alpha = 0.0024$), though several showed large uncorrected effect sizes against Llama: ChatGPT vs.\ Llama ($d=0.74$, $p_\text{uncorrected}=0.004$), Claude vs.\ Llama ($d=0.76$, $p_\text{uncorrected}=0.010$), DeepSeek vs.\ Llama ($d=0.79$, $p_\text{uncorrected}=0.011$). The significant omnibus test is thus driven primarily by Llama's lower satisfaction rather than differentiation among leaders.

\begin{figure}[t]
\centering
\includegraphics[width=\columnwidth]{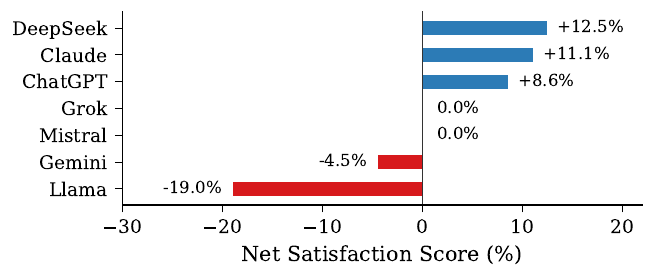}
\caption{Net Satisfaction Score by platform (NSS = \% Extremely satisfied $-$ \% Dissatisfied).}
\label{fig:nss}
\end{figure}

We also computed a Net Satisfaction Score (NSS), defined as the percentage of respondents reporting ``Extremely satisfied'' minus the percentage reporting either level of dissatisfaction. DeepSeek had the highest NSS (+12.5\%), followed by Claude (+11.1\%) and ChatGPT (+8.6\%). Grok and Mistral were neutral (0.0\% each). Gemini was the only major platform with a negative NSS ($-4.5\%$), despite having the third-largest user base; Llama scored $-19.0\%$ (Figure~\ref{fig:nss}). The NSS metric complements mean satisfaction by capturing the intensity of positive versus negative sentiment, which may better predict word-of-mouth recommendation behavior.

\begin{table}[t]
\centering
\caption{Satisfaction Summary Statistics.}
\label{tab:satisfaction}
\small
\setlength{\tabcolsep}{3pt}
\begin{tabular}{lrllrr}
\toprule
\textbf{Platform} & \textbf{n} & \textbf{M (SD)} & \textbf{95\% CI} & \textbf{NSS} & \textbf{\% Ext.} \\
\midrule
ChatGPT & 139 & 3.78 (1.10) & [3.60, 3.97] & +8.6 & 27.3 \\
Claude & 45 & 3.80 (1.01) & [3.50, 4.10] & +11.1 & 24.4 \\
Gemini & 88 & 3.51 (0.97) & [3.31, 3.71] & $-$4.5 & 12.5 \\
DeepSeek & 40 & 3.78 (0.86) & [3.51, 4.04] & +12.5 & 17.5 \\
Grok & 26 & 3.50 (0.95) & [3.14, 3.86] & 0.0 & 11.5 \\
Mistral & 19 & 3.32 (1.34) & [2.72, 3.92] & 0.0 & 21.1 \\
Llama & 21 & 2.95 (1.32) & [2.39, 3.52] & $-$19.0 & 14.3 \\
\bottomrule
\end{tabular}
\vspace{2pt}
\\\footnotesize\textit{Note:} NSS = Net Satisfaction Score (\%). Ext.\ = \% Extremely satisfied.
\end{table}

Free-tier and paid-tier users did not differ significantly in satisfaction for ChatGPT (free $M=3.72$ vs.\ paid $M=3.87$; $U=2178$, $p=0.358$), Claude ($p=0.241$), or Gemini ($p=0.536$), suggesting that satisfaction tracks the base model experience rather than premium feature access.

\noindent\textbf{Key takeaway:} The top three platforms are rated almost identically on satisfaction despite very different resources, histories, and business models. Paying for a subscription does not meaningfully increase satisfaction. For ChatGPT, longer tenure was positively associated with satisfaction (Spearman $\rho=0.18$, $p=0.036$, $n=139$), consistent with survivorship bias: less satisfied users are more likely to have discontinued use.

\begin{figure}[t]
\centering
\includegraphics[width=\columnwidth]{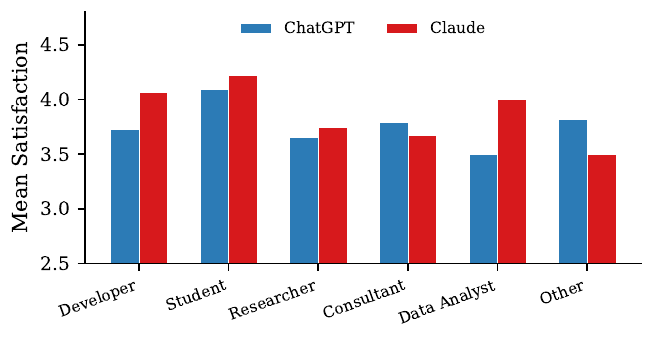}
\caption{Mean satisfaction for ChatGPT and Claude by occupation.}
\label{fig:occ_sat}
\end{figure}

\textbf{Satisfaction by professional role.} To examine whether professional context moderates platform preferences, we compared satisfaction ratings between the two largest occupational groups: developers/software engineers ($n=48$) and students ($n=38$). Among developers, Claude received higher mean satisfaction ($M=4.07$, $n=14$) than ChatGPT ($M=3.73$, $n=33$). Among students, both platforms were rated highly (Claude $M=4.22$, $n=9$; ChatGPT $M=4.09$, $n=32$). Researchers and consultants showed little platform differentiation. These patterns are consistent with Claude's stronger positioning among technical users, though the small per-cell sample sizes limit inferential confidence (Figure~\ref{fig:occ_sat}).

\subsection{Why Users Choose: Adoption Drivers and Domain Specialization}

\begin{figure}[t]
\centering
\includegraphics[width=\columnwidth]{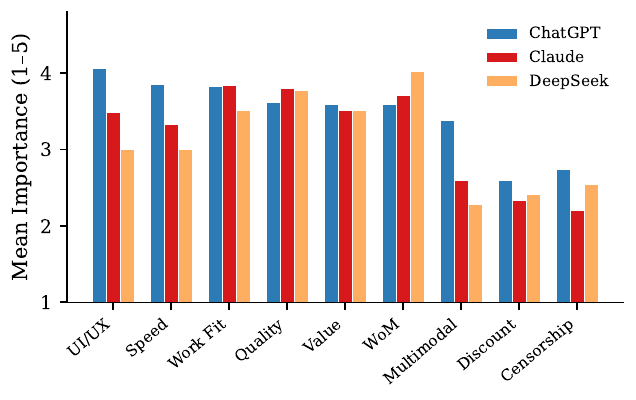}
\caption{Adoption driver importance for the three leading platforms rated on a 1--5 scale.}
\label{fig:drivers}
\end{figure}

If satisfaction scores converge at the top, what differentiates platforms in users' minds? The adoption driver data show that each leading platform attracts users for different reasons (Figure~\ref{fig:drivers}). ChatGPT's strongest driver was UI/UX design ($M=4.06$, $SD=1.08$, $n=139$), the highest single-driver score among leading platforms, followed by response speed ($M=3.86$) and work-task suitability ($M=3.83$). This profile suggests that among ChatGPT users, interface experience and workflow integration are rated as more important adoption factors than perceived output superiority.

Claude's adoption was driven by work-task suitability ($M=3.84$, $SD=1.31$, $n=45$), answer quality ($M=3.80$), and word-of-mouth ($M=3.71$). Claude's user base included a higher share of developers (31.1\% vs.\ 23.7\% for ChatGPT), though this difference was not statistically significant ($\chi^2=0.62$, $p=0.430$). Notably, Claude scored substantially lower on UI/UX ($M=3.47$) and speed ($M=3.76$) compared to ChatGPT, suggesting specific areas for product improvement.

DeepSeek exhibited the most distinctive adoption profile: word-of-mouth was its strongest driver ($M=4.02$, $SD=1.11$, $n=41$), the single highest driver score across all platform-driver combinations. This community-driven adoption, combined with low UI/UX ($M=3.00$) and speed ($M=3.00$) ratings, describes a platform that users find through social recommendation and tolerate despite interface limitations, primarily for its output quality.

\noindent\textbf{Key takeaway:} No single attribute drives adoption universally. Each platform attracts users for different reasons, and these distinct profiles create natural market segmentation.

Grok was uniquely characterized by censorship/content policy alignment as its top driver ($M=3.69$, $SD=1.23$, $n=26$), the highest censorship score across all platforms. Grok thus occupies a niche defined more by content policy than by capability, drawing users who find other platforms overly restrictive.

\begin{figure}[t]
\centering
\includegraphics[width=\columnwidth]{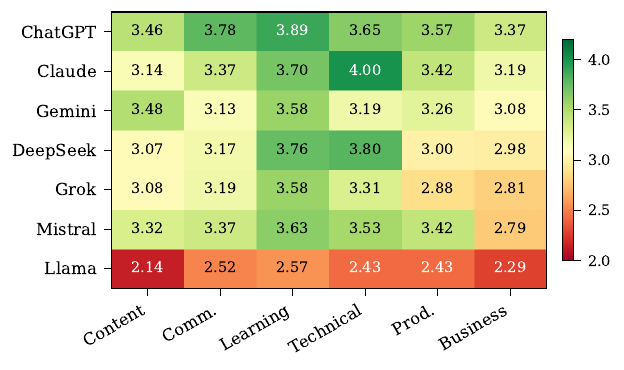}
\caption{Use case performance heatmap. Darker green indicates higher mean ratings; red indicates lower ratings.}
\label{fig:heatmap}
\end{figure}

Table~\ref{tab:usecase} and Figure~\ref{fig:heatmap} present mean use case ratings across all platform-category combinations, further illustrating this specialization.

Claude achieved the highest absolute score of any platform in any category for Technical and Analytical Tasks ($M=4.00$, $SD=1.11$, $n=43$), significantly higher than the next competitor, DeepSeek ($M=3.80$), and well above Gemini ($M=3.19$). ChatGPT led in Learning and Research ($M=3.89$, $SD=1.03$, $n=139$) and Communication Assistance ($M=3.78$), reflecting its strength as a generalist tool. Gemini showed relative strength in Content Creation ($M=3.48$), slightly edging ChatGPT ($M=3.46$), possibly due to its multimodal capabilities and integration with Google's content tools.

Llama scored below the neutral midpoint across all six categories (range: $M=2.14$ to $M=2.57$), a finding that likely reflects the technical complexity of deploying and using open-source models rather than intrinsic model quality. Most Llama users in our sample accessed it through third-party interfaces rather than direct deployment, which may introduce additional friction and quality variation.

\noindent\textbf{Key takeaway:} Platforms have developed measurable domain strengths. Claude leads in technical tasks, ChatGPT in learning and communication, and Gemini in content creation.

\begin{table}[t]
\centering
\caption{Mean Use Case Performance Ratings (1--5 scale).}
\label{tab:usecase}
\small
\setlength{\tabcolsep}{3pt}
\begin{tabular}{lcccccc}
\toprule
 & \textbf{Cont.} & \textbf{Comm.} & \textbf{Learn.} & \textbf{Tech.} & \textbf{Prod.} & \textbf{Bus.} \\
\midrule
ChatGPT & 3.46 & \textbf{3.78} & \textbf{3.89} & 3.65 & \textbf{3.57} & \textbf{3.37} \\
Claude & 3.14 & 3.37 & 3.70 & \textbf{4.00} & 3.42 & 3.19 \\
Gemini & \textbf{3.48} & 3.13 & 3.58 & 3.19 & 3.26 & 3.08 \\
DeepSeek & 3.07 & 3.17 & 3.76 & 3.80 & 3.00 & 2.98 \\
Grok & 3.08 & 3.19 & 3.58 & 3.31 & 2.88 & 2.81 \\
Mistral & 3.32 & 3.37 & 3.63 & 3.53 & 3.42 & 2.79 \\
Llama & 2.14 & 2.52 & 2.57 & 2.43 & 2.43 & 2.29 \\
\bottomrule
\end{tabular}
\vspace{2pt}
\\\footnotesize\textit{Note:} Bold = highest in category. Cont. = Content Creation; Comm. = Communication; Learn. = Learning \& Research; Tech. = Technical; Prod. = Productivity; Bus. = Business.
\end{table}

A Kruskal-Wallis test on the Technical Tasks dimension across the four largest platforms (ChatGPT, Claude, Gemini, DeepSeek) found a significant difference ($H=12.84$, $df=3$, $p=0.005$). Post-hoc comparisons revealed that Claude's Technical score significantly exceeded Gemini's ($U=1056$, $p=0.001$, $d=0.77$), while the Claude-DeepSeek and Claude-ChatGPT comparisons were not significant after correction ($p=0.28$ and $p=0.07$ respectively).

\subsection{First-Mover Advantage, Tenure, and Anchoring Effects}

\begin{figure}[t]
\centering
\includegraphics[width=\columnwidth]{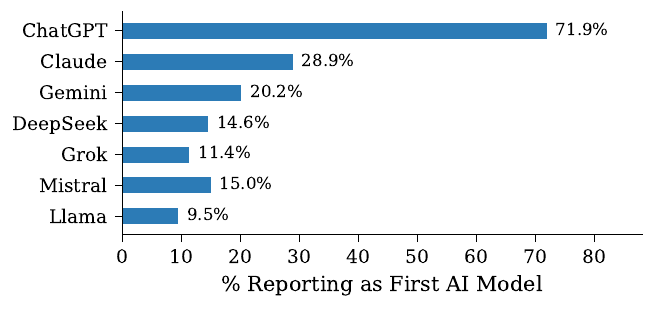}
\caption{Percentage of each platform's users who report it as their first AI chat model.}
\label{fig:firstmodel}
\end{figure}

The specialization patterns above describe why users might prefer different platforms, but they do not explain how users arrived at their current choices. ChatGPT serves as the dominant entry point to the AI chat ecosystem: 71.9\% of its users (100/139) reported it as their first AI chat model, far exceeding Claude (28.9\%), Mistral (30.0\%), Gemini (20.2\%), DeepSeek (14.6\%), Llama (9.5\%), and Grok (3.8\%) (Figure~\ref{fig:firstmodel}). This first-mover advantage has clear implications for habit formation and default effects.

\begin{figure}[t]
\centering
\includegraphics[width=\columnwidth]{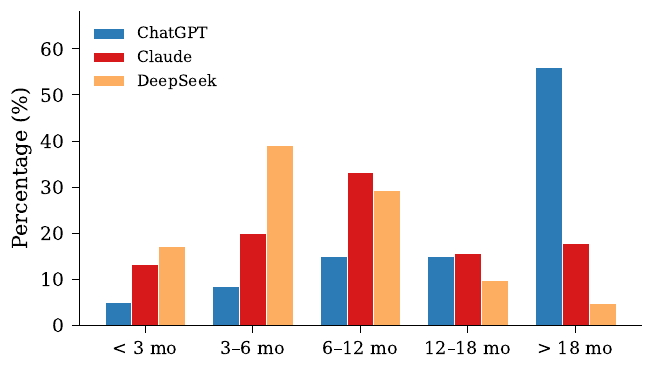}
\caption{Usage tenure distribution for the three leading platforms.}
\label{fig:tenure}
\end{figure}

The tenure distribution further reinforces ChatGPT's entrenchment (Figure~\ref{fig:tenure}). Among ChatGPT users, 56.1\% have used the platform for over 18 months, reflecting adoption shortly after its November 2022 launch. By contrast, 53.7\% of DeepSeek users have been on the platform for less than 6 months, consistent with its breakout visibility following the DeepSeek-R1 release in early 2025. Claude occupies an intermediate position, with 33.3\% in the 6--12 month range and 17.8\% exceeding 18 months, reflecting steadier growth anchored in developer communities.

These tenure patterns have implications for satisfaction measurement: ChatGPT's longer tenure may introduce survivorship bias (dissatisfied users have already left), while newer platforms may benefit from novelty effects. We return to this issue in Section~\ref{sec:discussion}.

\noindent\textbf{Key takeaway:} ChatGPT's market position is anchored in early adoption and long tenure rather than superior satisfaction, while newer entrants like DeepSeek are growing rapidly through a different user base.

\begin{figure}[t]
\centering
\includegraphics[width=\columnwidth]{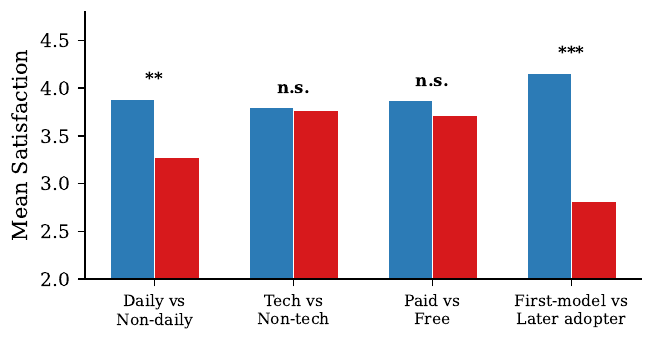}
\caption{ChatGPT satisfaction by subgroup. Significance: ***$p<0.001$, **$p<0.01$, n.s.\ = not significant. All tests are Mann-Whitney $U$, uncorrected.}
\label{fig:subgroups}
\end{figure}

To test whether first-mover status translates into a satisfaction advantage, we conducted exploratory subgroup comparisons within the ChatGPT user base ($n=139$), our largest platform sample (Figure~\ref{fig:subgroups}). These are reported as descriptive patterns rather than confirmatory tests; $p$-values are uncorrected for multiplicity.

\textbf{Usage frequency.} Daily ChatGPT users reported higher satisfaction ($M=3.89$, $n=114$) than non-daily users ($M=3.28$, $n=25$; $U=992$, $p=0.003$). This may reflect genuine quality appreciation among heavy users, but could also reflect self-selection: users who find the tool unsatisfying may naturally reduce usage frequency.

\textbf{First-model status.} The largest observed difference was between respondents who reported ChatGPT as their first AI model ($M=4.16$, $n=100$) and those who adopted it after trying another platform ($M=2.82$, $n=39$; $U=942$, $p<0.001$). This 1.34-point gap is striking and consistent with an anchoring or status-quo effect: users whose initial AI experience was ChatGPT rate it substantially higher than those who came to it with prior expectations set by a competitor. This finding reinforces the importance of first-mover advantage not only for market share but for perceived satisfaction.

\textbf{Subscription tier.} Paid ChatGPT users ($M=3.87$, $n=69$) and free users ($M=3.72$, $n=69$) did not differ significantly ($p=0.358$), suggesting that satisfaction is shaped by the base model experience rather than premium features.

\textbf{Professional background.} Technical users (developers, data scientists, researchers; $n=61$) and non-technical users ($n=78$) reported similar ChatGPT satisfaction ($M=3.80$ vs.\ $M=3.77$, $p=0.998$). For Claude, technical users trended higher ($M=3.96$, $n=23$) than non-technical users ($M=3.64$, $n=22$; $p=0.403$), though this did not reach significance.

\subsection{Pricing and Price Sensitivity}

Beyond satisfaction and switching dynamics, pricing structure shapes which platforms users can access and how they respond to cost changes. ChatGPT has the highest paid penetration among surveyed platforms, with 38.9\% of respondents on individual paid plans and an additional 9.4\% on team or enterprise plans (Table~\ref{tab:plans}). Claude shows a similar paid profile (30.4\% individual), while DeepSeek operates with 92.7\% free-tier users and Llama with 90.5\% free, reflecting their open-source or freemium positioning.

\begin{table}[t]
\centering
\caption{Subscription Plan Distribution.}
\label{tab:plans}
\small
\setlength{\tabcolsep}{3pt}
\begin{tabular}{lllr}
\toprule
\textbf{Platform} & \textbf{Free} & \textbf{Paid Indiv.} & \textbf{n} \\
\midrule
ChatGPT & 76 (51.0\%) & 58 (38.9\%) & 149 \\
Claude & 26 (56.5\%) & 14 (30.4\%) & 46 \\
Gemini & 56 (62.2\%) & 18 (20.0\%) & 90 \\
DeepSeek & 38 (92.7\%) & 1 (2.4\%) & 41 \\
Grok & 21 (77.8\%) & 4 (14.8\%) & 27 \\
Mistral & 10 (50.0\%) & 7 (35.0\%) & 20 \\
Llama & 19 (90.5\%) & 0 (0.0\%) & 21 \\
\bottomrule
\end{tabular}
\vspace{2pt}
\\\footnotesize\textit{Note:} Team/Enterprise and ``Unsure'' omitted for brevity.
\end{table}

\begin{figure}[t]
\centering
\includegraphics[width=\columnwidth]{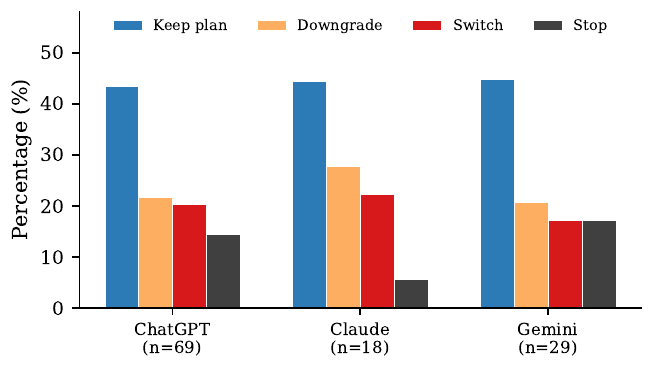}
\caption{Response to a hypothetical 25\% price increase among paid subscribers.}
\label{fig:price}
\end{figure}

To gauge price elasticity, we asked paid subscribers how they would respond to a hypothetical 25\% price increase. Among paid ChatGPT users ($n=69$): 43.5\% would keep their current plan, 21.7\% would downgrade, 20.3\% would switch to a competitor, and 14.5\% would stop using AI tools entirely. Claude paid users ($n=18$) showed slightly stronger retention: 44.4\% keep, 27.8\% downgrade, 22.2\% switch, and only 5.6\% stop (Figure~\ref{fig:price}). A chi-square test found no significant cross-platform differences in price response distributions ($\chi^2=2.37$, $df=6$, $p=0.883$, Cram\'{e}r's $V=0.10$). This suggests that price sensitivity is a market-wide phenomenon rather than platform-specific, with roughly 55--60\% of paid users across all platforms expressing willingness to change behavior after a moderate price increase.

\noindent\textbf{Key takeaway:} Price sensitivity is high and platform-agnostic. Free alternatives exert continuous downward pressure on premium tiers across all providers.

\subsection{User Frustrations and Feature Gaps}

The quantitative results above characterize what users choose and how satisfied they are; the open-ended responses reveal what they still find lacking. A total of 329 open-ended responses were collected across two prompts: frustrations ($n=170$) and desired features ($n=159$). We present these as descriptive thematic summaries rather than formal qualitative findings, since coding was performed by the research team without independent inter-rater reliability assessment. Initial codes were generated inductively from the first 50 frustration responses; these were then applied to the full corpus with iterative refinement. Counts should be read as rough salience indicators rather than prevalence estimates. Table~\ref{tab:frustrations} presents the ten most frequently occurring themes.

\begin{table}[t]
\centering
\caption{User Frustrations by Theme ($n=170$ responses).}
\label{tab:frustrations}
\small
\setlength{\tabcolsep}{4pt}
\begin{tabular}{lrrl}
\toprule
\textbf{Theme} & \textbf{n} & \textbf{\%} & \textbf{Illustrative Examples} \\
\midrule
Hallucination & 38 & 22.4 & Fabricated citations, facts \\
Censorship & 26 & 15.3 & Over-refusals, creative blocks \\
Context loss & 17 & 10.0 & Forgetting prior instructions \\
Usage limits & 15 & 8.8 & Free-tier message caps \\
Slow responses & 14 & 8.2 & Latency on premium tiers \\
Instruction fail & 13 & 7.6 & Ignoring format directives \\
Verbosity & 13 & 7.6 & Excessive preambles, lists \\
Repetition & 10 & 5.9 & Looping, recycled outputs \\
Poor reasoning & 8 & 4.7 & Arithmetic, logic errors \\
Sycophancy & 5 & 2.9 & Agreeing rather than correcting \\
\bottomrule
\end{tabular}
\end{table}

The top two frustrations, hallucination and censorship, are instructive because they represent opposing failure modes. Hallucination reflects insufficient constraint (the model generates plausible but false content), while censorship reflects excessive constraint (the model refuses legitimate requests or sanitizes creative output). The tradeoff is direct: tightening filters to reduce hallucination risk may increase perceived censorship, and vice versa. Users are sensitive to both failure modes, placing platforms in a difficult optimization space.

Frustration patterns varied across platforms. Censorship complaints were disproportionately directed at ChatGPT and Claude, while hallucination was cited most frequently for ChatGPT and Gemini. Context loss was most commonly cited for ChatGPT, possibly reflecting its heavier use for longer conversations. Sycophancy complaints targeted Claude specifically, suggesting that its conversational style may sometimes prioritize agreeableness over correctness.

\begin{figure}[t]
\centering
\includegraphics[width=\columnwidth]{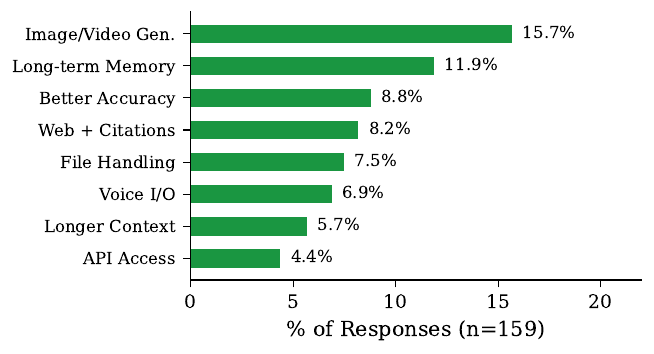}
\caption{Most desired features across all platforms ($n=159$ responses).}
\label{fig:features}
\end{figure}

Analysis of desired features (Figure~\ref{fig:features}) revealed that image and video generation was the most requested capability (15.7\%, 25/159), followed by long-term memory across conversations (11.9\%, 19/159), better factual accuracy (8.8\%, 14/159), and web access with source citations (8.2\%, 13/159). The demand for multimodal generation and persistent memory indicates that users already conceive of AI assistants as general-purpose creative and organizational tools, well beyond text-based question answering.

\noindent\textbf{Key takeaway:} Users' top frustrations (hallucination and censorship) represent a direct engineering tradeoff. Their top feature requests (multimodal generation and persistent memory) signal where the market is heading.

\section{Discussion}
\label{sec:discussion}

\subsection{Competitive Plurality Through Specialization}

The most consequential finding for industry observers is the satisfaction convergence among leading platforms. Claude, ChatGPT, and DeepSeek all fall between 3.78 and 3.80 on a five-point scale, with confidence intervals that overlap substantially. Although we cannot rule out that this convergence partly reflects scale compression on a five-point measure, the pattern is consistent across multiple metrics (mean satisfaction, NSS, and use case ratings), suggesting it reflects a genuine narrowing of perceived quality differences. While the omnibus Kruskal-Wallis test is significant ($p=0.012$), this is driven by Llama's lower scores (Cohen's $d = 0.74$ to $0.79$ vs.\ the top tier) rather than differences among leaders. This convergence is notable given the resource asymmetries involved: OpenAI has raised over \$50 billion and employs thousands of researchers, while DeepSeek emerged from a Chinese quantitative hedge fund with a fraction of that compute budget, and Anthropic operates with substantially fewer resources than Google's Gemini project.

Combined with pervasive multi-homing (over four in five respondents use two or more platforms, and more than half use three or more), this challenges, at least among the active users in our sample, the winner-take-all predictions that have dominated industry analysis. Unlike social networks where lock-in derives from network effects~\cite{katz1985}, or mobile platforms where switching costs are high due to app ecosystems~\cite{zhu2012}, AI chat tools function as commoditized utilities where users freely mix and match. The negligible switching costs (users can simply open a different browser tab) mean that no platform can rely on structural lock-in to retain users.

Although aggregate satisfaction scores converge, the use case data reveal specialization that plausibly sustains coexistence. Claude's leadership in Technical Tasks ($M=4.00$), combined with its developer-heavy user base (31.1\% developers) and quality-driven adoption profile, is consistent with an audience concentrated among technical professionals. ChatGPT's strength in Learning ($M=3.89$), Communication ($M=3.78$), and Productivity ($M=3.57$), paired with UI/UX-driven adoption ($M=4.06$), is consistent with a generalist positioning sustained by experience polish and ecosystem breadth rather than raw output quality among this user population.

DeepSeek represents a third competitive model: technically competitive output ($M=3.80$ Technical, $M=3.76$ Learning) with community-driven adoption (word-of-mouth $M=4.02$) and near-universal free access (92.7\% free users). Its growth trajectory (53.7\% of users joined in the past six months) suggests a platform in the rapid-expansion phase, where reputation precedes product maturity. This tripartite structure mirrors Cennamo and Santalo's~\cite{cennamo2013} framework of platform competition through differentiation: as markets mature, platforms compete less on breadth and more on domain-specific excellence.

\subsection{First-Mover Advantage, Anchoring, and the Persistence of Defaults}

Our tenure analysis reveals an important asymmetry in how users experience different platforms. ChatGPT's user base is dominated by long-tenure users (56.1\% over 18 months), while DeepSeek's base is predominantly recent adopters (53.7\% under 6 months). This creates at least two measurement challenges. First, ChatGPT's satisfaction scores may reflect survivorship bias: users who were dissatisfied in early interactions may have already departed, leaving behind a self-selected population of satisfied users. Second, DeepSeek's scores may benefit from a novelty effect, where the excitement of discovering a new, capable, and free tool inflates initial satisfaction. The near-identical satisfaction means ($M=3.78$ for both) may thus represent different points on each platform's satisfaction trajectory rather than equivalent steady-state evaluations.

Our exploratory subgroup analysis reinforces this interpretation. ChatGPT users who reported it as their first AI model rated satisfaction 1.34 points higher than those who adopted it after trying another platform ($M=4.16$ vs.\ $M=2.82$, $p<0.001$). While this cross-sectional comparison cannot establish causality, the magnitude of the difference is consistent with strong anchoring effects: users whose expectations were set by ChatGPT rate it far more favorably than users who arrived with prior experience of alternatives.

The first-mover data also highlights the persistence of defaults in technology adoption. With 71.9\% of ChatGPT users reporting it as their first AI model, ChatGPT may benefit from a form of status quo bias, where users continue with a familiar option despite available alternatives. Users who begin with ChatGPT may persist out of habit rather than active preference. This reading is consistent with our finding that among ChatGPT users, UI/UX received the highest mean importance rating ($M=4.06$) while output quality scored lower ($M=3.55$), suggesting that interface familiarity may be a more salient factor in continued use than answer quality per se. If confirmed longitudinally, this pattern would imply that the window for displacing ChatGPT as the default consumer AI tool is narrowing with each month of habit formation, even as competitors match or exceed its raw capabilities.

\subsection{Product Tradeoffs: Hallucination, Censorship, and Pricing Pressure}

The two most prevalent frustrations, hallucination (22.4\%) and content filtering (15.3\%), represent a standing tension in LLM product design. Reducing hallucination risk typically requires more conservative generation, stronger refusal behaviors, and hedged outputs, all of which increase the likelihood of triggering censorship complaints. Conversely, reducing perceived censorship by allowing more open-ended generation increases the surface area for hallucination.

Grok's unique censorship-aligned positioning ($M=3.69$ censorship driver, highest across all platforms) reveals a genuine market segment of users who prioritize content permissiveness. However, Grok's middling satisfaction ($M=3.50$, NSS = 0.0\%) suggests that content policy alignment alone is insufficient to drive high overall satisfaction; it must be paired with competitive capability. No platform has resolved this tradeoff to universal satisfaction, and it may prove to be a durable axis of market segmentation.

The finding that 55--60\% of paid users across all platforms would modify their behavior in response to a 25\% price increase indicates non-trivial price elasticity. The availability of high-quality free alternatives, particularly DeepSeek ($M=3.78$ satisfaction, 92.7\% free) and the improving Llama ecosystem, exerts continuous downward pressure on premium pricing. For providers seeking to sustain premium pricing, these data suggest that differentiation through integration depth, reliability, enterprise features, and trust may matter more than base model quality alone.

\subsection{Implications for LLM Evaluation Practice}

The data point to a gap between benchmark rankings and user satisfaction. Despite frequent leaderboard position changes and heated debates about which model leads on specific benchmarks, the top three consumer platforms achieve statistically indistinguishable user satisfaction. Once models cross a capability threshold that the current leaders appear to have reached, interface quality, response style, content policy, pricing, and ecosystem integration may matter more to users than marginal performance gains on automated tests.

Supplementing capability-focused benchmarks with user-centric metrics would help capture the platform experience as users actually encounter it. Our six-item use case scale (Cronbach's $\alpha$ = 0.79--0.85) and nine-item adoption driver scale ($\alpha$ = 0.75--0.80) offer starting points, but the field needs standardized instruments that can be administered across platforms and over time. Without such instruments, the evaluation community risks optimizing for dimensions of quality that matter less to users than interface design, content policy, or ecosystem integration.

\section{Limitations}
\label{sec:limitations}

Several limitations constrain the generalizability of our findings, and we have framed our claims accordingly as characterizing active, engaged AI chat users rather than the broader consumer population. First, the sample substantially overrepresents power users: 79.5\% report daily usage, and technology professionals (developers, data scientists, researchers) constitute a majority. Casual users, who may have different satisfaction patterns and adoption drivers, are underrepresented. Our robustness checks (Section 3.5) show that satisfaction estimates are stable across full completers and partial responders and across early and late collection periods, but they cannot rule out selection effects relative to the broader population. Second, while respondents spanned 37 countries, geographic concentration in North America (38.2\%) and South Asia (36.4\%) limits claims about global preferences; users in East Asia, Africa, and Latin America are particularly underrepresented. Third, small sample sizes for Mistral ($n=19$) and Llama ($n=21$) constrain statistical power for these platforms; findings involving them should be treated as descriptive. Fourth, self-selection bias is inherent in voluntary online surveys: respondents motivated to share opinions about AI tools may differ systematically from the broader user population. Fifth, the within-subjects evaluation design, while enabling direct cross-platform comparison, may introduce anchoring effects whereby ratings for one platform influence subsequent ratings. Sixth, thematic coding of open-ended responses was performed by the research team without formal inter-rater reliability assessment; we present qualitative results as descriptive summaries (Section 4.6) accordingly. Seventh, this is a cross-sectional study capturing a single snapshot; longitudinal data would be needed to establish causal relationships. Eighth, the mid-survey addition of DeepSeek and Mistral as explicit model options (Section 3.1) introduces a potential instrument effect, though early ``Other'' coding and temporal stability checks suggest this did not materially alter response patterns. Ninth, the use of a five-point satisfaction scale may introduce ceiling compression: if most users are at least moderately satisfied, mean scores will cluster in the 3.5--4.0 range regardless of genuine preference differences. The satisfaction convergence we report should be interpreted with this measurement property in mind; finer-grained scales or comparative forced-choice designs might reveal preference differences that our instrument could not detect.

\section{Future Work}

Several directions for future work emerge from this study's findings and limitations.

\textbf{Larger and more representative samples.} Future studies should employ stratified probability sampling to reach casual users, non-English speakers, and populations outside technology-focused communities. Pre-registered hypothesis testing with larger platform-specific samples would allow confirmatory analysis of the exploratory patterns identified here.

\textbf{Longitudinal tracking.} A panel design following users over 6--12 months could disentangle survivorship bias from genuine satisfaction, track how users respond to major model updates, and measure whether the satisfaction convergence we observe is stable or transient. Conjoint analysis administered at multiple time points could rigorously quantify how users trade off attributes like speed, accuracy, price, and content policy.

\textbf{Social media and behavioral data.} Our planned next phase will supplement survey data with large-scale analysis of public discourse on platforms such as Reddit and X (formerly Twitter), where users frequently compare AI tools, report switching decisions, and describe frustrations in natural language. Computational methods including sentiment analysis, topic modeling, and stance detection applied to AI-related subreddits and hashtags could capture preference signals at much larger scale and with less self-selection bias than voluntary surveys. Combining these observational data with behavioral usage logs (collected with informed consent) would allow triangulation of self-reported preferences against revealed behavior.

\textbf{Benchmark--satisfaction alignment.} A direct investigation of whether benchmark improvements translate into measurable user experience gains would be valuable. Correlating Chatbot Arena Elo rating changes with longitudinal satisfaction data from the same user population could establish whether the gap between benchmarks and user perception is narrowing or widening as models improve.

\textbf{Cross-cultural comparison.} The distinct regional compositions of platform user bases (e.g., DeepSeek's strong adoption in South and East Asia, ChatGPT's dominance in North America) may reflect cultural preferences in communication style, content expectations, and interface design. A multi-language, multi-region study with culturally adapted instruments would test whether the satisfaction convergence and multi-homing patterns we observe generalize beyond English-speaking, technology-oriented communities.

\textbf{Enterprise and API users.} Our sample captures consumer-facing chat usage but not the growing enterprise and developer segment, where API pricing, latency, and tool integration may reshape the preference landscape. Industry surveys already suggest different competitive dynamics in enterprise contexts~\cite{menlo2025}; dedicated research on this segment would complement consumer-focused studies like ours.

\section{Ethical Considerations}

All participation was voluntary and anonymous. No personally identifiable information was collected beyond approximate geographic location (country of residence). Respondents were informed of the study's purpose and could withdraw at any time without consequence. The study did not involve vulnerable populations, deception, or sensitive personal questions beyond technology usage patterns. Informed consent was obtained at the beginning of the survey.

We note that our survey included questions about commercial products and services, which carries a risk of implicit endorsement or criticism. We have attempted to present findings neutrally, reporting satisfaction differences as empirical observations rather than product recommendations. The research team has no financial relationships with any of the AI platform companies discussed in this study. We also acknowledge that distributing surveys through technology-focused communities may inadvertently amplify the voices of early adopters at the expense of broader population perspectives, and we encourage future researchers to employ more inclusive sampling strategies.

\section{Conclusion}

This study provides an early cross-platform empirical baseline for understanding how active AI chat users evaluate competing platforms. Surveying 388 respondents across seven platforms, we find reported satisfaction parity among leading platforms (though this may partly reflect measurement compression on a five-point scale), pervasive multi-model usage (82.4\% use two or more), distinct adoption profiles, meaningful domain specialization, moderate price sensitivity, and persistent frustrations centered on hallucination and content moderation.

Among the active, technology-oriented users in our sample, the overarching pattern is competitive plurality rather than market consolidation. These users assemble portfolios of AI assistants tailored to different tasks and preferences rather than converging on a single winner. ChatGPT retains its first-mover advantage (71.9\% of its users report it as their first AI model, and 56.1\% have used it for over 18 months), but this entrenchment does not translate to satisfaction dominance. Claude, DeepSeek, and others have established viable competitive positions by excelling in domains valued by their respective user segments.

For platform developers, durable differentiation will depend on the full user experience, including interface design, response quality, content policy calibration, and ecosystem integration, rather than chasing marginal benchmark improvements. The distinct adoption profiles we identify offer actionable guidance: ChatGPT's competitive position rests on its UI/UX lead, Claude's on technical excellence among developers, DeepSeek's on community-driven adoption and free access, and Grok's on content-policy alignment. Each faces distinct challenges: ChatGPT must justify premium pricing against free alternatives, Claude must expand beyond its technical niche, DeepSeek must improve interface quality to retain users acquired through word-of-mouth, and Grok must pair its policy stance with competitive capability.

For evaluation researchers, user-centric assessment complements automated benchmarks by measuring whether daily users are satisfied with what they receive. The satisfaction convergence at the top, despite benchmark variation, suggests that above a quality threshold non-capability factors dominate, and standardized instruments for measuring multi-dimensional platform experience (building on scales like our six-item use case measure) would be a useful addition to the evaluation toolkit.

For policymakers, the competitive dynamics we observe among active users (low switching costs, widespread multi-homing, and viable alternatives to the market leader) suggest that at least for engaged, tech-savvy consumers, the AI assistant market offers meaningful choice. However, the concentration of first-mover advantage (71.9\% entry through ChatGPT) and the emerging role of ecosystem lock-in through API integrations and enterprise contracts warrant continued monitoring as the market matures.

\bibliographystyle{plain}

\end{document}